# A Fixed-Volume Variant of Gibbs-Ensemble Monte Carlo Yields Significant Speedup in Binodal Calculation

Sanbo Qin[1] and Huan-Xiang Zhou[1,2,a]

[1]Department of Chemistry and [2]Department of Physics, University of Illinois Chicago, Chicago, IL 60607

[a]Correspondence: hzhou43@uic.edu

**Abstract**

Gibbs-ensemble Monte Carlo (GEMC) is a powerful method for calculating the gas-liquid binodals of simple models and small molecules, but is too demanding computationally for realistic models of proteins. Here we discover that the main reason for long simulations is that volume exchange is very slow to achieve, and develop a variant GEMC without volume exchange. The key is to determine an appropriate initial density. Test of this fixed-volume GEMC method on Lennard-Jones and patchy particles shows enormous speedup without any loss of accuracy in predicted binodals. The fast speed of fixed-volume GEMC promises many applications.

I.   **Introduction**

Since the introduction of the Gibbs-ensemble Monte Carlo (GEMC) method by Panagiotopoulos in 1987,[1] it has been widely used to calculate the binodals of vapor-liquid and liquid-liquid phase separation for various fluids, including simple models such as Lennard-Jones and patchy particles, small molecules such as water and hydrocarbons, and homopolymers of various lengths as well as mixtures.[2-18] In particular, based on GEMC simulations of patchy particle mixtures, the relative strength of intermolecular interactions was identified as a key determinant for how a macromolecular component can regulate the phase equilibrium of biomolecular condensates.[12, 13] Several books or book chapters and reviews provide comprehensive coverage on the implementation and applications of GEMC.[19-22]

GEMC involves two boxes, initially with the same volume and the same particle number. The simulation proceeds with particle exchange between the two boxes to achieve equality in chemical potential and volume exchange to achieve equality in pressure, along with particle translation/rotation to achieve equilibration within each box. Particle exchange and particle translation/rotation are order $N$ operations, but volume exchange is an order $N^2$ operation. Here $N$ denotes the total particle number. In a typical simulation, each Monte Carlo cycle consists of $N$ particle-exchange attempts, $N$ particle-translation/rotation attempts, and 5 volume-exchange attempts. The final densities in the two boxes represent the coexistence densities of the gas (or dilute) and liquid (or dense) phases. A plot of these densities as a function of temperature is known as a binodal.

Many variations of the original GEMC have been introduced, in particular to gain speedup. For mixtures, Panagiotopoulos et al.[23] presented a "constant-pressure" version, in which the two boxes, instead of exchanging volumes with each other, change their volumes independently at a given pressure. For large molecules such as long polymers, particle exchange is very difficult as the chance of a large molecule being accepted into a dense box is extremely small. In some cases, polymer chains are restricted to the dense-phase box.[10, 11] Different ideas have been proposed to facilitate the exchange of large molecules. For example, instead of transferring a polymer chain from one box to another in a single step, a tagged chain is transferred monomer by monomer.[3] For a mixture of large and small molecules, a large molecule in one box can be exchanged with multiple small molecules in the other box.[14, 17] Recently, field-theoretic simulations of polymers have been used to implement the Gibbs ensemble, allowing thorough sampling of chain configurations and exchange between the boxes [24].

In the present work, we observed that, in a typical GEMC simulation, particle exchange occurs very early, but volume exchange takes at least 100 times longer. Based on this observation, we present a fixed-volume variant of GEMC, where the two boxes undergo particle exchange but not volume exchange. By choosing an appropriate initial density, we obtain the correct binodal but with an enormous reduction in simulation time compared to the standard GEMC. The outline of the rest of the paper is as follows. Section II demonstrates a significant lag of volume exchange behind particle exchange. In Section III, we show that fixed-volume GEMC can produce the correct binodal if an appropriate initial density is chosen, and present a method for determining this initial density and the resulting coexistence densities. Section IV illustrates the speedup and accuracy of our fixed-volume GEMC method when benchmarked against the standard GEMC. We conclude the paper with some remarks in Section V.

## II. Volume exchange lags particle exchange in GEMC

We first report some results from simulations using the standard GEMC. We chose two types of particles. The first is Lennard-Jones particles, with interactions given by the well-known Lennard-Jones potential. We will use the contact distance ($\sigma$) as the unit of length. We apply a distance cutoff of 3 and shift the energy function so it becomes 0 at the cutoff distance. The second is patchy particles of Kern and Frenkel,[5] each with four attractive patches on the surface of a sphere, situated at the vertices of a tetrahedron and together covering 70% of the spherical surface. Here we use the diameter of the sphere as the unit of length. The range of attraction is 1.5. Temperature (denoted by $T$) will be in units of $\varepsilon/k_B$, where $\varepsilon$ is the well depth and $k_B$ is the Boltzmann constant.

In Fig. 1, we present traces of particle numbers, volumes, and densities in the two boxes from GEMC simulations of Lennard-Jones particles. Each Monte Carlo cycle consists of $N$ particle-exchange attempts, $N$ particle-translation attempts, and 5 volume-exchange attempts. Particle numbers reach a steady level after only $10^2$ to $5 \times 10^2$ cycles (Fig. 1a-c), but volumes take $10^4$ to $5 \times 10^5$ cycles to reach a steady level (Fig. 1d-f). After volumes reach their steady levels, particle numbers adjust to a new steady level. Correspondingly, densities display two steady levels, one before and one after volumes reach steady levels (Fig. 1g-i). As the temperature nears the critical value, the box that contains more particles frequently changes from one to the other.

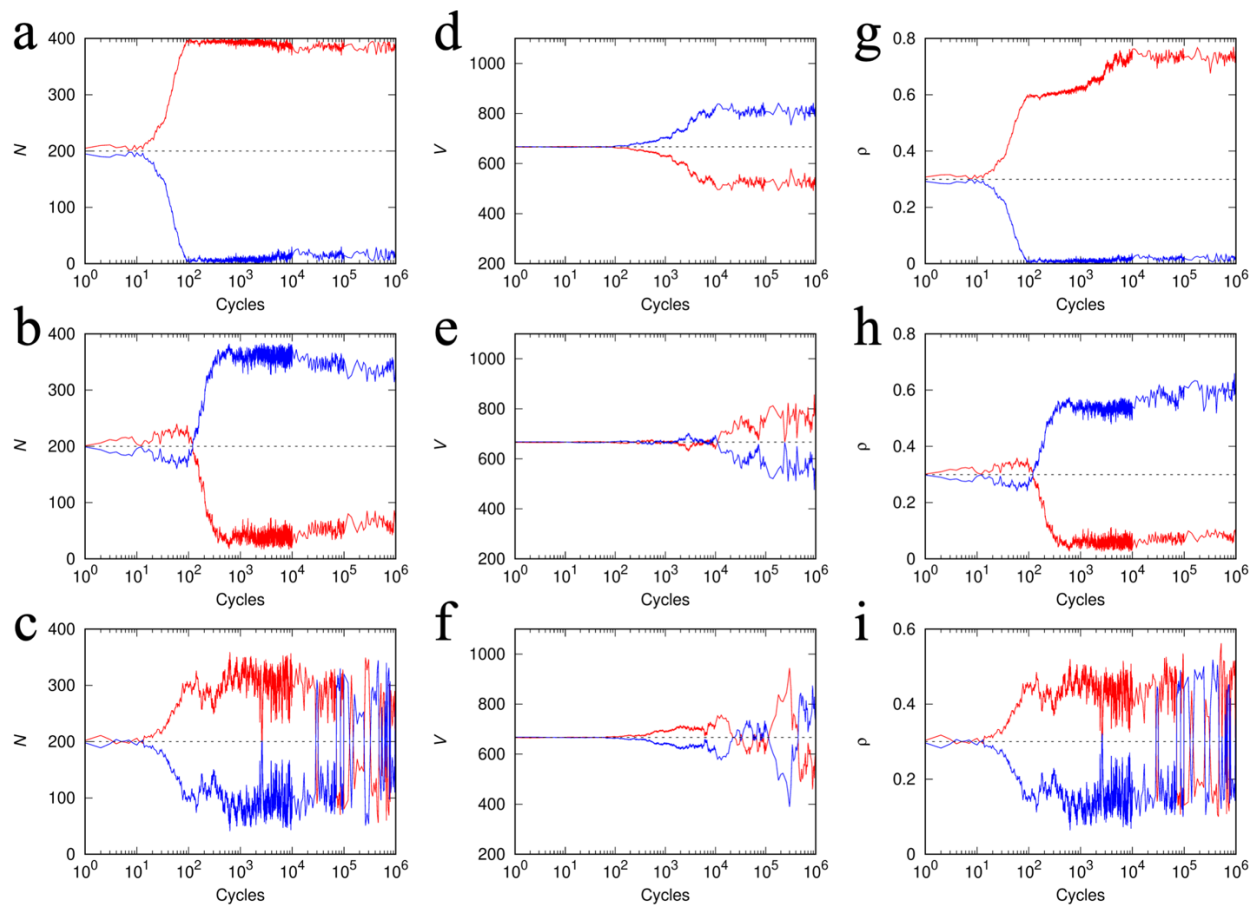

**FIG. 1.** Particle numbers, volumes, and densities in the two boxes in GEMC simulations of 400 Lennard-Jones particles. (a-c) Particles number at $T = 0.85$, 1.05, and 1.15. (d-f) Volumes at $T = 0.85$, 1.05, and 1.15. (g-h) Densities at $T = 0.85$, 1.05, and 1.15. To reduce clutter, only every 10th, 1000th, and 20000th points are used for plotting when the cycle count are between $10^2$ and $10^4$, between $10^4$ and $10^5$, and $> 10^5$, respectively.

GEMC simulations of patchy particles exhibit very similar behaviors (Fig. 2). Here a particle that is selected for movement is either translated or rotated (split with even chances). Particle numbers reach a steady level after ~$10^3$ cycles, but volumes reach a steady level only after ~$10^5$ cycles. As the temperature approaches the critical value, the box that represents the liquid phase (with higher density) frequently changes hands. This high frequency arises partly from the fact that the more populous box changes hands as noted for Lennard-Jones particles, but also simply from fluctuations in particle numbers and volumes. We note that frequent phase switches can result in errors in the coexistence densities, as densities in the two boxes move toward the mean density during the switches. Consequently, the gas-phase density would be inflated and the liquid-phase density would be suppressed.

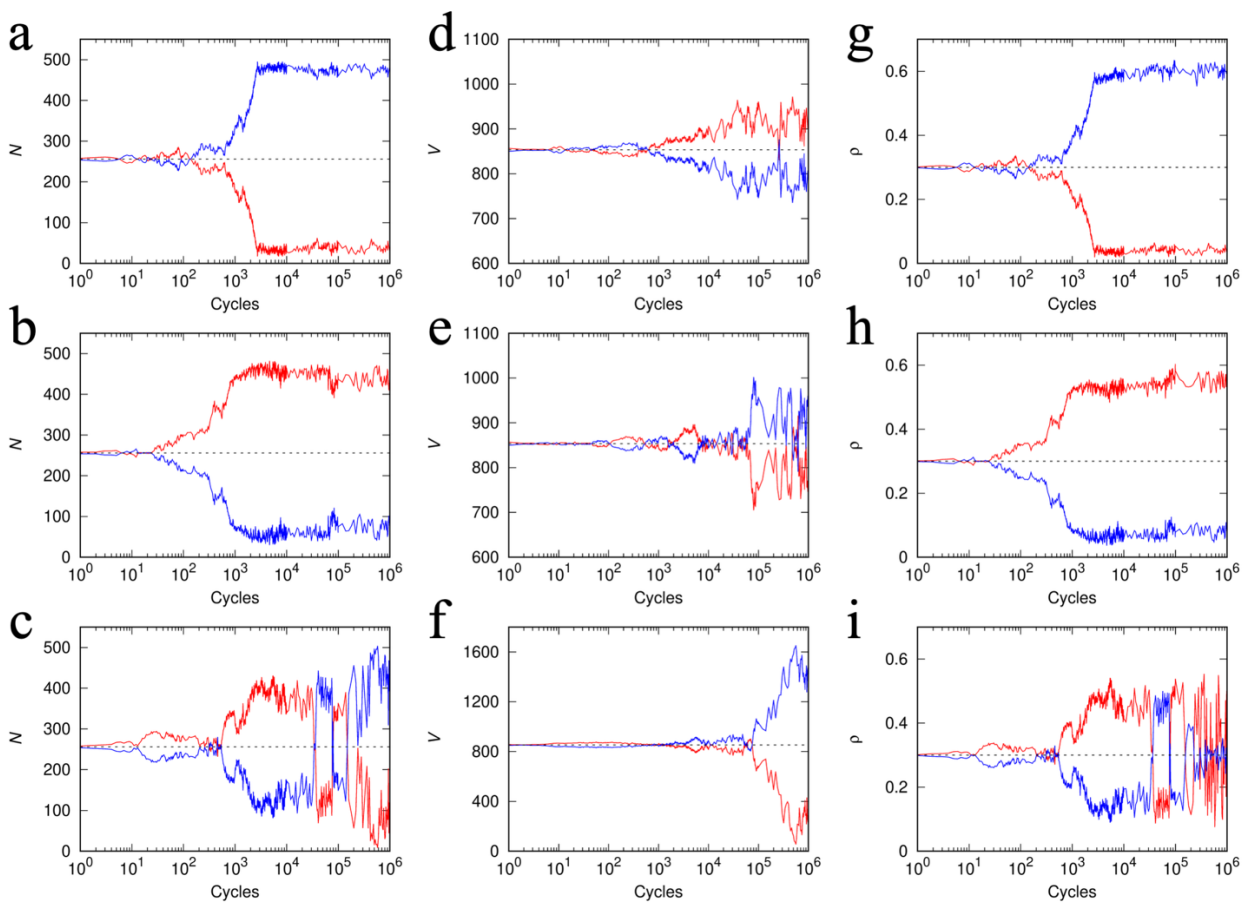

**FIG. 2.** Particle numbers, volumes, and densities in the two boxes in GEMC simulations of 512 patchy particles. (a-c) Particles number at $T = 0.66$, 0.70, and 0.74. (d-f) Volumes at $T = 0.66$, 0.70, and 0.74. (g-h) Densities at $T = 0.66$, 0.70, and 0.74. The initial density is 0.3. Every point, every 10th, 1000th, and 20000th points are used for plotting when the cycle count are $< 10^2$, between $10^2$ and $10^4$, between $10^4$ and $10^5$, and $\geq 10^5$, respectively. Horizontal dashed lines indicate initial values.

The fact that it takes at least 100 times more cycles to achieve volume exchange than particle exchange raises the question of whether we can eliminate volume exchange in order to speed up GEMC simulations. Next we show that that is indeed possible.

### III. Fixed-volume GEMC can produce the correct binodal

### 1. The standard GEMC

In the standard GEMC, particle exchange ensures equality in chemical potential and volume exchange ensures pressure equality. Of course both boxes are at the same temperature. The total particle number ($N$) and the total volume ($V$) are conserved in the simulation. After reaching phase equilibrium, let the densities in the two boxes be $\rho_1$ and $\rho_2$. The foregoing four quantities allow the determination of the particle numbers, $N_1$ and $N_2$, in and volumes, $V_1$ and $V_2$, of the two equilibrated boxes. We have

$$N_1 + N_2 = N, \qquad [1]$$

$$V_1 + V_2 = V, \qquad [2]$$

$$N_1/V_1 = \rho_1, \qquad [3]$$

$$N_2/V_2 = \rho_2. \qquad [4]$$

Solving these equations, we find

$$V_1 = -\frac{N - \rho_2 V}{\rho_2 - \rho_1}, \qquad [5]$$

$$V_2 = \frac{N - \rho_1 V}{\rho_2 - \rho_1}. \qquad [6]$$

Note that we can choose any initial density,

$$\rho_0 \equiv \frac{N}{V}, \qquad [7]$$

inside the binodal. The simulation will always produce the same equilibrium densities $\rho_1$ and $\rho_2$ in the two boxes. The freedom in choosing initial densities will be crucial for the variant GEMC to be presented next.

The equilibrium densities, $\rho_1$ and $\rho_2$, are determined by equality in chemical potential and pressure between the phases. They can be found by a Maxwell equal-area construction on the $\mu - \rho$ plane, assuming a constant temperature, as illustrated in Fig. 3a, where a red horizontal line bisects the $\mu - \rho$ isotherm in a way that ensures equality in areas above and below and hence equality in pressure. The isotherm has a nonmonotonic segment known as the van der Waals loop, which occurs under temperature and pressure conditions where two phases coexist.

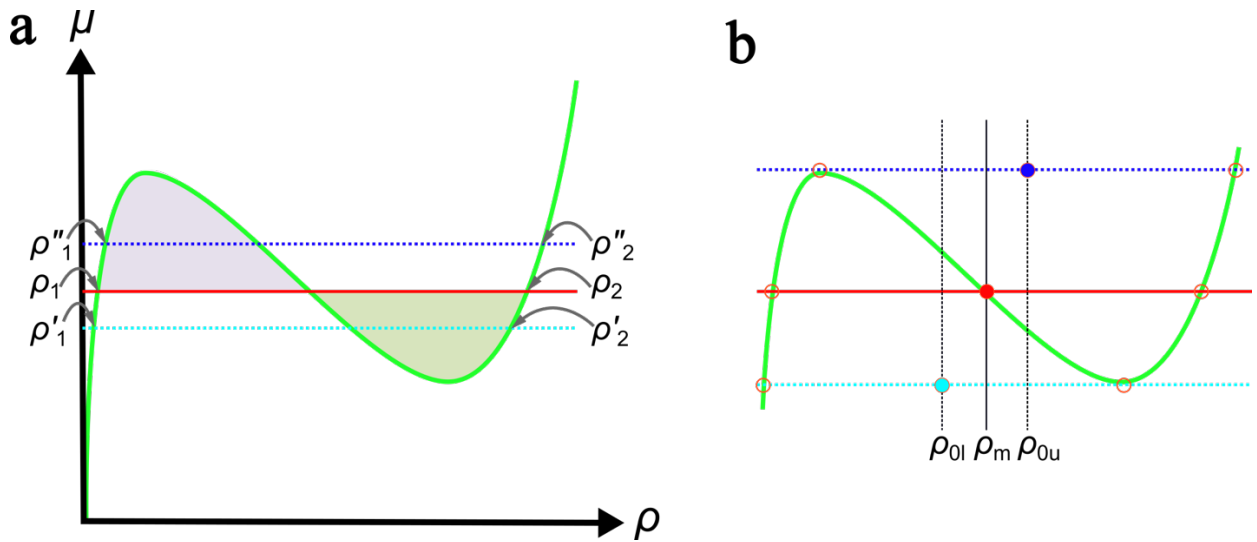

**FIG. 3.** Determination of the densities in the two phases, according to the standard or fixed-volume GEMC. (a) Horizontal lines that bisect the $\mu - \rho$ isotherm (green curve) correspond to fixed-volume GEMC, which only ensures equality in chemical potential. The red horizontal line is special, as it bisects the $\mu - \rho$ isotherm with equal areas above (violet shading) and below (olive shading), thereby ensuring equality in pressure. The resulting $\rho_1$ and $\rho_2$ are the densities at phase equilibrium, which would be obtained by the standard GEMC. In comparison, the cyan horizontal line leads to lower pressure in the dense phase than in the light phase, and densities ($\rho'_1$ and $\rho'_2$) that are lower than their phase equilibrium counterparts. Conversely, the blue horizontal line leads to densities ($\rho''_1$ and $\rho''_2$) that are higher than their phase equilibrium counterparts. (b) Initial densities that allow fixed-volume GEMC to both achieve equality in chemical potential and satisfy the constraint of Eq. [12] are limited to a narrow range, from $\rho_{0l}$ to

$\rho_{0u}$. In addition, $\rho_m$, the mean density of the two equilibrated phases, is near the midpoint of $\rho_{0l}$ and $\rho_{0u}$.

## 2. Fixed-volume GEMC

Given the significant lag of volume exchange (Figs. 1 and 2), we want to eliminate volume exchange in GEMC. That is, we start the two boxes at the same volume,

$$V_1 = V_2 = V/2, \qquad [8]$$

as the standard GEMC method, but then keep the volumes fixed at these initial values. We can reach the same phase equilibrium as the standard GEMC if we choose the right initial density. We just need to ensure that the equilibrium values for $V_1$ and $V_2$, given by Eqs. [5] and [6], satisfy Eq. [8]. That is,

$$\frac{N - \rho_2 V}{\rho_1 - \rho_2} = -\frac{N - \rho_1 V}{\rho_1 - \rho_2}, \qquad [9]$$

which yields

$$V = \frac{N}{(\rho_1 + \rho_2)/2} \qquad [10]$$

and correspondingly,

$$\rho_0 \equiv \frac{N}{V}$$

$$= \frac{\rho_1 + \rho_2}{2} \equiv \rho_m. \qquad [11]$$

That is, if we pick the initial density to be the mean density, $\rho_m$, of the two equilibrated phases, then we will achieve the correct phase equilibrium!

For any other choice ($\rho'_0$) of initial density, fixed-volume GEMC leads to densities (denoted as $\rho'_1$ and $\rho'_2$) that are different from the equilibrium values ($\rho_1$ and $\rho_2$). Fixed-volume GEMC is equivalent to a Maxwell construction without the equal-area requirement (any of the horizontal lines in Fig. 3a). Instead of equal area, $\rho'_1$ and $\rho'_2$ are constrained, based on Eq. [8], by

$$\rho'_1 + \rho'_2 = 2\rho'_0. \tag{12}$$

One immediate consequence of Eq. [12] is that the density in the dense phase cannot be higher than $2\rho'_0$.

Fixed-volume GEMC must meet two requirements: achieving equality in chemical potential and satisfying the constraint of Eq. [12]. A crucial observation is that initial densities that allow fixed-volume GEMC to meet the twin requirements are limited to a narrow range (FIG. 3b). It is bracketed from below by $\rho_{0l}$, which corresponds to the cyan horizontal line that is tangent to the lower tip of the van der Waals loop of the $\mu - \rho$ isotherm, and from above by $\rho_{0u}$, which corresponds to the blue horizontal line that is tangent to the upper tip of the van der Waals loop. The van der Waals loop typically has relatively small amplitudes, which decrease with increasing particle numbers.[25] Of course the desired initial density, $\rho_m$, falls inside the range $[\rho_{0l}, \rho_{0u}]$. Therefore, the desired initial density can be bracketed in a narrow range.

Another important observation is that $\rho_m$ is not only inside $[\rho_{0l}, \rho_{0u}]$ but should be very close to the midpoint of this range. Because $\rho_m$ corresponds to the line (red in Fig. 3b) that bisects the van der Waals loop with equal areas below and above, the two lines (cyan and blue in Fig. 3b) that are tangent to the lower and upper tips should lie on the opposite sides of the equal-area line with approximately equal distances. In short, if we can find $\rho_{0l}$ and $\rho_{0u}$, then their midpoint gives a good estimate of $\rho_m$.

## 3. Determination of $\rho_{0l}$ and $\rho_{0u}$

Note that inside the van der Waals loop, the $\mu - \rho$ isotherm has positive slopes in both the left and right portions (Fig. 3). The positivity of the slopes is dictated by the positivity of the isothermal compressibility of physical systems, which are expected to contract upon increasing pressure. Correspondingly, when the initial density $\rho'_0$ is within $[\rho_{0l}, \rho_{0u}]$, the densities, $\rho'_1$ and $\rho'_2$, in the two phases produced by fixed-volume GEMC are both growing functions of $\mu$. Their mean value, $\rho'_0$, is also a growing function of $\mu$. Treating $\mu$ as an intermediate variable, we can see that $\rho'_1$ and $\rho'_2$ are both growing functions of $\rho'_0$ when the latter is within $[\rho_{0l}, \rho_{0u}]$.

Figure 4 displays the dependences of $\rho'_1$ and $\rho'_2$ on $\rho'_0$ from fix-volume GEMC simulations of Lennard-Jones particles at $T = 0.95$. Nine simulations are run at 9 initial densities ranging from 0.1 to 0.5 in increments of 0.05. We place $\rho'_1$ and $\rho'_2$ on the $x$-axis and $\rho'_0$ on the $y$-axis, to facilitate comparison with Fig. 3. We notice that the $\rho'_1$ curve has negative slopes at low $\rho'_0$ and the $\rho'_2$ curve has negative slopes at high $\rho'_0$. Only when $\rho'_0$ is within a narrow range in the middle do both the $\rho'_1$ and $\rho'_2$ curves have positive slopes. This is the range, $[\rho_{0l}, \rho_{0u}]$, we are looking for. Specifically, $\rho_{0l}$ is the initial density at which the slope of $\rho'_1$ with respect to $\rho'_0$ is 0, and $\rho_{0u}$ is the initial density at which the slope of $\rho'_2$ with respect to $\rho'_0$ is 0. Equivalently, $\rho'_1$ is at its minimum when $\rho'_0 = \rho_{0l}$ and $\rho'_2$ is at its maximum when $\rho'_0 = \rho_{0u}$.

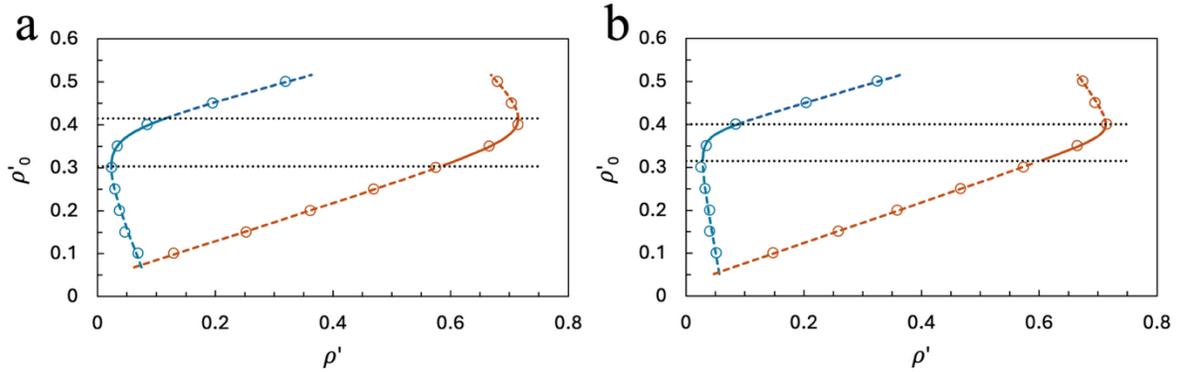

**FIG. 4.** Dependences of fixed-volume GEMC $\rho'_1$ and $\rho'_2$ on the initial density $\rho'_0$ for Lennard-Jones particles at $T = 0.95$. (a) $N = 400$. (b) $N = 1200$. Circles are simulation results; curves are hyperbolic fits. The two horizontal lines display $\rho'_0 = \rho_{0l}$ and $\rho'_0 = \rho_{0u}$. The hyperbolic curves are in solid when inside these two lines and in dash when outside.

To precisely determine $\rho_{0l}$ and $\rho_{0u}$, we fit the $\rho'_1$ curve to an equation. We notice that these curves have a hyperbolic shape, and choose the equation to be that for a hyperbola. For the present purpose, it is most convenient to use a parametric representation of the hyperbolic equation:

$$\mathbf{x} = \mathbf{s}_0 + \mathbf{s}_1 t + \mathbf{s}_2 \frac{1}{t}, \qquad [13]$$

where bold symbols represent vectors in the *xy* plane, $\mathbf{x}$ is any point on the hyperbola, $\mathbf{s}_0$ denotes the center of the hyperbola, $\mathbf{s}_1$ and $\mathbf{s}_2$ are along the asymptotes of the hyperbola. Once the fitting parameters ($\mathbf{s}_0$, $\mathbf{s}_1$, and $\mathbf{s}_2$) are determined by the $\rho'_1$ data, we use Eq. [12] to obtain the hyperbolic equation for the $\rho'_2$ curve. The resulting hyperbolas for Lennard-Jones particles at $T = 0.95$ are shown in Fig. 4 as blue curves for $\rho'_1$ and brown curves for $\rho'_2$. The segments for $\rho'_0$ within $[\rho_{0l}, \rho_{0u}]$ are in solid whereas those outside this range are in dash. $\rho_{0l}$ and $\rho_{0u}$ are obtained from the zero-slope conditions of the $\rho'_1$ and $\rho'_2$ hyperbolas. We display $\rho'_0 = \rho_{0l}$ and $\rho'_0 = \rho_{0u}$ as horizontal lines. The results at $N = 400$ are presented in Fig. 4a whereas those at N = 1200 are in Fig. 4b. The results at the two system sizes are very close to each other, but, as expected, the $[\rho_{0l}, \rho_{0u}]$ range shows a slight narrowing at the larger system size.

## 4. Predicting the correct binodal

We now have a method for using fixed-volume GEMC to predict binodals. The simulation part consists of fixed-volume GEMC runs at 5 to 10 initial densities spanning a certain range. The postprocessing part consists of fitting the resulting dependence of $\rho'_1$ on $\rho'_0$ to a hyperbola, and using the resulting hyperbolic equation to calculate $\rho_{0l}$ and $\rho_{0u}$. Lastly, we use the mean value of $\rho_{0l}$ and $\rho_{0u}$ as the initial density to predict $\rho_1$ and $\rho_2$ from the hyperbolic equation.

## IV. Speedup and accuracy of fixed-volume GEMC

We display in Fig. 5a-c the particle numbers and the corresponding densities in fixed-volume GEMC simulations of 400 Lennard-Jones particles at $T = 0.85$, 1.05, and 1.15. Particle numbers in the two boxes reach steady levels by $\sim 5 \times 10^2$ cycles. Each cycle consists of $N$ particle-translation and $N$ particle-exchange attempts. The Monte Carlo cycles required and the steady levels reached are similar to those in the standard GEMC simulations before significant volume exchange takes place (Fig. 1a-c), as expected. After particle numbers in the two boxes break from each other and reach steady levels, there is little chance for the two boxes to switch phases. Similar observations can be made for fixed-volume GEMC simulations of 512 patchy particles at $T = 0.66$, 0.70, and 0.74, except that the Monte Carlo cycles required for particle numbers to reach steady levels are increased to ~$5 \times 10^3$ (Fig. 5d-f). For both types of particles, we choose to run fixed-volume GEMC for $10^4$ cycles and use the last $5 \times 10^3$ for calculating densities in the two boxes.

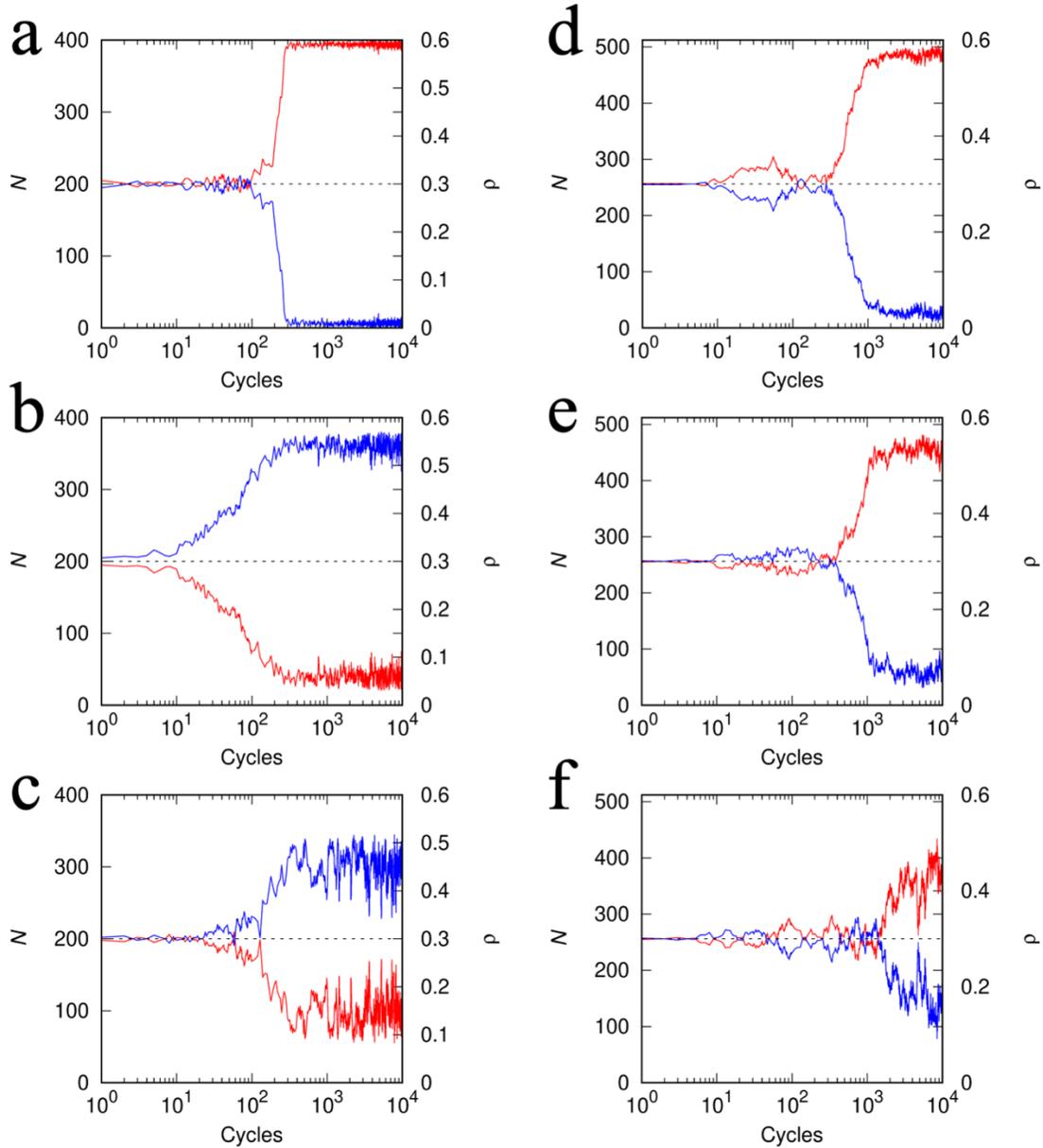

**FIG. 5.** Particle numbers (left ordinate) and densities (right ordinate) in the two boxes in fixed-volume GEMC simulations. (a-c) 400 Lennard-Jones particles at $T = 0.85$, 1.01, and 1.15. (d-f) 512 patchy particles at $T = 0.66$, 0.70, and 0.74. Every point and every 10th point are used for plotting when the cycle count are $< 10^2$ and $\geq 10^2$, respectively.

For $10^4$ cycles, a fixed-volume GEMC simulation of 400 Lennard-Jones particles takes 32.3 s to run on a single thread of an Intel Xeon E5-2650 v4 2.20 GHz processor. In comparison, the standard GEMC, with 5 volume-exchange attempts along with $N$ particle-translation and $N$ particle-exchange attempts per cycle, takes 58.9 s. The near doubling of CPU time is due to the fact that volume exchange is an order $N^2$ operation ($N^2/2$ to be more precise), whereas particle translation and particle exchange are order $N$ operations. The expected ratio of CPU times is thus $(5 \times \frac{1}{2} + 1 + 1) / (1 + 1) = 2.5$, close to the observed doubling of CPU time. Since the standard GEMC needs to run 100 times longer ($10^6$ cycles instead of $10^4$), the speedup of our fixed-volume GEMC method is $2 \times 100 / 9 \approx 22$, where the number 9 accounts for the number of initial densities at which we run fixed-volume GEMC simulations.

Lastly, in Fig. 6 we compare the coexistence densities from the standard GEMC simulations for $10^6$ cycles with those from fixed-volume GEMC simulations for $10^4$ cycles. The agreement overall is excellent. For Lennard-Jones particles (Fig. 6a), fixed-volume GEMC slightly underestimates the densities at the lowest temperature, $T = 0.65$. On the other hand, at the highest temperature, $T = 1.15$, the GEMC densities are clearly too low, as indicated by an apparent departure of the mean density from linear dependence on temperature, as demanded by the law of rectilinear diameter [26]. In contrast, the fixed-volume GEMC mean densities follow a linear dependence on temperature for all temperatures. For the standard GEMC, the simulations at initial densities from 0.1 to 0.5 all give consistent results for the coexistence densities at $T$ up to 0.95; the range of "good" initial densities narrows to [0.15, 0.45] at $T = 1.05$ and further to [0.25, 0.35] at $T = 1.15$.

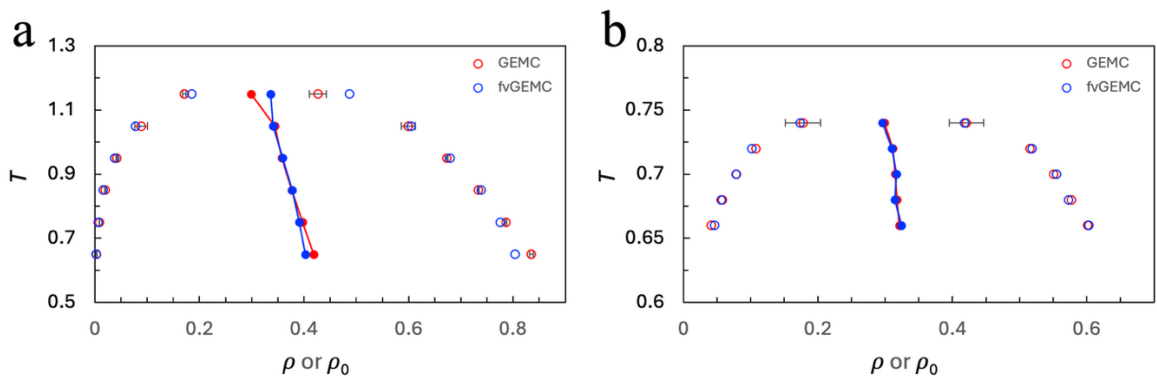

**FIG. 6.** Comparison of binodals from standard and fixed-volume GEMC. (a) 400 Lennard-Jones particles. For GEMC, results are averaged over those from simulations at 3-9 initial densities in the range of 0.1 to 0.5; error bars represent standard deviations. (b) 512 patchy particles. For GEMC, results are from a single GEMC simulation at an initial density of 0.3, except for those at $T = 0.74$, which are the averages and standard deviations (displayed as error bars) of 11 replicate simulations. Open circles are coexistence densities, and closed circles are their mean values. Standard and fixed-volume GEMC simulations have $10^6$ and $10^4$ cycles, respectively. Densities are averaged over the second half of the simulations, always taking the lower and higher density values from the two boxes for the gas and liquid phases, respectively.

Binodal comparison between standard and fixed-volume GEMC for patchy particles is presented in Fig. 6b. Here the two methods give essentially identical results for all temperatures. We note an interesting observation from the standard GEMC at the highest temperature, $T = 0.74$. In 14 replicate simulations at an initial density of 0.3, three end at a catastrophe, where the first box is reduced to essentially zero volume and the second box contains all the particles. In the second box, particles are not distributed uniformly (Fig. 7). Instead, the top of the box has large voids while the bottom has particles densely packed. It appears as if the second box contains two phases. The densities of the two boxes would be calculated as 0 and 0.3, respectively. For reporting the coexistence densities, we exclude the three catastrophic simulations when calculating the mean and standard deviation. Had we included them, the densities of both phases would be lowered, leading to a departure of the mean density from linear dependence on temperature. The departure of the mean density at $T = 1.15$ from linearity in the binodal of Lennard-Jones particles likely can also be attributed to the occurrence of mixed phases in the two boxes.

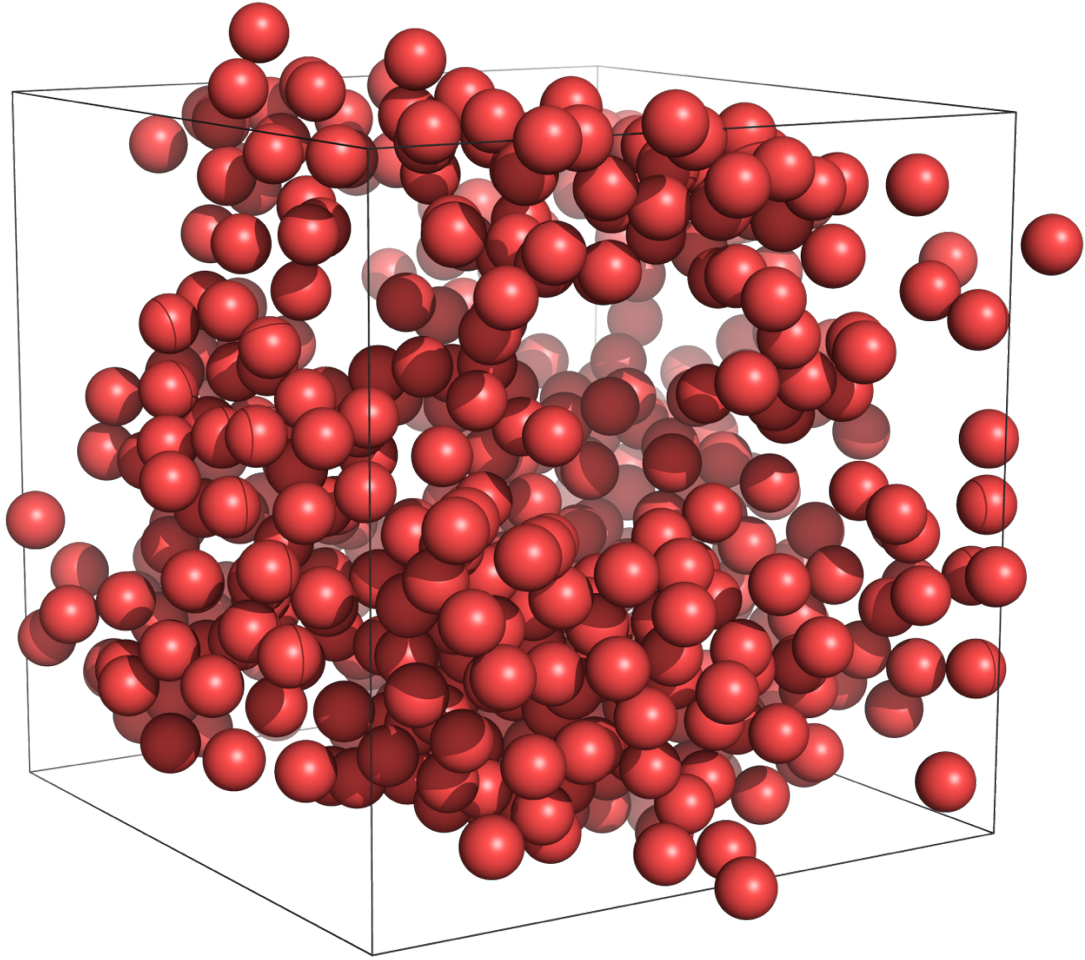

**FIG. 7.** Final snapshot of a GEMC simulation of patchy particles at $T = 0.74$. One box has 0 volume; the snapshot shown is for the second box, which contains all the 512 particles. The top and bottom of the box appear to represent the gas and liquid phases, respectively.

## V. Conclusion

We have developed a fixed-volume GEMC method that gains enormous speedup over the standard GEMC without any loss in accuracy. In contrast to the standard GEMC, fixed-volume GEMC has little chance for the two boxes to switch phases. Frequent phase switches can result in errors in the coexistence densities, as the densities in the two boxes move toward the mean density during the switches. Moreover, as the temperature approaches the critical value, volume exchange can lead to a catastrophe, where one box is reduced to 0 volume and the other box contains mixed phases.

Our fixed-volume GEMC method requires simulations at 5 to 10 initial densities. Because these simulations are independent, they can easily run in parallel on multiple cores or threads. In contrast, parallelization of GEMC codes is complicated and hardly ever achieves linear scaling.[22] While the initial densities by design cover a relatively wide range, only simulations within a narrow range correspond to or are close to phase equilibrium. Outside this narrow range, the simulated systems violate the positivity of isothermal compressibility (Fig. 4) and hence are not thermodynamically stable. Instead of clean single phases, the two boxes contain mixed phases, similar to the situation in Fig. 7. Extension to mixtures poses an interesting problem.

We anticipate many applications of the fixed-volume GEMC method. Its fast speed makes it well-suited for high-throughput or iterative use, such as for force-field refinements.[8, 15, 18] It may also be feasible for use on realistic (e.g., all-atom) models of proteins,[27] potentially leading to residue-level accuracy in predicted binodals.

## Acknowledgement

This work was supported by National Institutes of Health Grant GM118091.